# NEW NANOSTRUCTURES OF CARBON: QUASI-FULLERENES $C_{n-q}$ (n= 20, 42, 48, 60)

Christian A. Celaya[1], Jesús Muñiz[2,3], Luis Enrique Sansores[1*]

[1]Departamento de Materiales de Baja Dimensionalidad, Instituto de Investigaciones en Materiales, Universidad Nacional Autónoma de México, Apartado, Postal 70-360, Ciudad de México, 04510, México.

[2]Instituto de Energías Renovables, Universidad Nacional Autónoma de México, Priv. Xochicalco s/n, Col. Centro, Temixco, Morelos CP 62580, México

[3]CONACYT-Universidad Nacional Autónoma de México, Priv. Xochicalco s/n, Col. Centro, Temixco, Morelos CP 62580, México

**Abstract**

Based on the third allotropic form of carbon (Fullerenes) through theoretical study have been predicted structures described as non-classical fullerenes. We have studied novel allotropic carbon structures with a closed cage configuration that have been predicted for the first time, by using DFT at the B3LYP level. Such carbon $C_{n-q}$ structures (where, n=20, 42, 48 and 60), combine states of hybridization $sp^1$ and $sp^2$, for the formation of bonds. A comparative analysis of quasi-fullerenes with respect to their isomers of greater stability was also performed. Chemical stability was evaluated with the criteria of aromaticity through the different rings that build the systems. The results show new isomerism of carbon nanostructures with interesting chemical properties such as hardness, chemical potential and HOMO-LUMO gaps. We also studied thermal stability with Lagrangian molecular dynamics method using Atom- Center Density propagation (ADMP) method.

Keywords: DFT, quasi-fullerenes, allotropic, encapsulation, energy storage.

## 1. Introduction

Since the discovery of $C_{60}$ fullerene [1], known as the third allotropic form of carbon [2,3]. The closed-cage carbon structures have been of great interest because of the rich chemical variety of compounds that can be formed [3,4]. The scientific community has focused its attention on them, due to the different type of fullerenes that can be formed (endofullerenes, metallofullerenes, quasi-fullerenes, among other.) [5]. In the search for new fullerenes, they have been able to synthesize a great variety in different sizes. A general classification is by its size into lower fullerenes ($C_n$ with n < 60) and higher fullerenes ($C_n$ with n > 60), which have the highest yield in the synthesis [6,3,5]. It is important to remark that the different isomers that fullerenes can adopt represent a great number of possible geometries that follows the isolated-pentagon rule (IPR) [6,7].

In general, fullerenes are defined as polyhedral closed cages made up entirely of *n* three-coordinate carbon atoms and having 12 pentagonal and (n/2-10) hexagonal faces, where n ≥ 20 [7] and satisfy the IPR. These structures are cataloged in allotropes of 0 Dimension [1–4,8,9]. In recent decades several methods have been reported, exploring the formation of fullerenes, where the most stable structures are those that satisfy the IPR [6,10,11]. Nevertheless, the IPR is not commonly satisfied with the small fullerenes, due to the possibility to form a great variety of isomers, since the small differences in formation energies of each isomer make difficult to identify [11-16]. Therefore, theoretical studies of all possible isomers that can be form with small fullerenes have been performed using molecular dynamics [17, 18], semi-empirical [14, 18, 24] and DFT methods [12, 13, 15, 16, 20–27].

Carbon is an allotropic element and external agents such as temperature, pressure, light, electric currents, modify its different structural forms and structural defects. Consequently, the variations of these factors improve the formation of new allotropic forms of carbon [6,11]. On other hand, novel methods have been reported during the synthesis of large fullerenes in addition to the formation of small carbon nanostructures in the process of discharge arc. This method is the most widely used for the formation of small carbon clusters [28, 29, 30]. In recent years, Kharlamov et al. [28] reported a novel method known as Flow Through Continuous Pyrolysis (FTCP), an alternative technique to the discharge arc for synthesis of fullerene $C_{60}$ and new carbon molecules, through the fullerenization of molecules of benzenes, which achieves temperatures around 1000 ºC. Mass spectra results, obtained by the method FTCP, showed the formation of fullerenes $C_{60}$, quasi-fullerenes ($C_{48}$, $C_{42}$ and $C_{40}$) and small clusters of carbon in gas phase [28,31].

On the other hand, the synthesis of small carbon structures opens a new field of research in the study of new allotropic forms of carbon that have not been reported [17–24,32]. In general polyhedral closed cages made up entirely of n three-coordinate carbon atoms shall be known as quasi-fullerenes [7]. As stated above, quasi-fullerenes may present rings with pentagonal or hexagonal symmetry. These kinds of systems do not follow the IPR, since they may contain four-membered, six-membered and eight-membered rings [17,23]. One of the technical difficulties that currently exist in the synthesis of quasi-fullerenes is their isolation. Quasi-fullerenes are smaller than fullerenes, resulting in low-energy stability [29,31]. In this respect, the study at the theoretical level in this class of molecules is essential to understand the chemical properties that the systems may exhibit.

A large number of theoretical works emphasizing the study of stability and formation of the different isomers that quasi-fullerenes may form have previously been performed [32-41]. The first studies [23,25-27,42-46] were performed using DFT and semiempirical [24,27,32,43] methods in order to find stabilized geometries of non-classical fullerenes with interesting electronic structure properties, due to the presence of different carbon rings. One novel method employed to find other forms of allotropic carbon is through molecular dynamics methods, where implemented algorithms allow the search of new quasi-fullerenes with regular and semi-regular polyhedral shapes [24]. Molecular Dynamics represents another choice to find other forms of allotropic carbon [17,18,24], where structures of regular shapes and semi-regular polyhedral have been found. In recent years, Belenkova and Shahova [32] reported some structures of carbon using molecular mechanics method (MM+). They found some attractive geometry that combined $sp^1$-$sp^2$-$sp^3$ hybridization states to form closed cage structures. These geometries have different rings in their structure (squares, hexagons, octagons). The result shows that the specific binding energy decreases as the fullerene radius decreases and as the fraction of sp$^2$ atoms increases. The structures most stable are those containing a maximum number of sp$^2$ hybridize atoms.

On the other hand, the varied molecular topography of quasi-fullerenes could have interesting properties and potential application for energy storage, encapsulation of transition metal and electronics devices. Despite of the large variety of theoretical works published about this type of molecules, it has not been predicted in detail their electronic and chemical properties at the DFT level. In this work, we analyzed some structures of carbon clusters, namely, $C_{20}$, $C_{42}$, $C_{48}$ and $C_{60}$, which combine states of hybridization and present different ring members. From the DFT level we studied their electronic structure, chemical stability, aromaticity, HOMO-LUMO gaps, chemical reactivity and chemical hardness. Thermal stability was also studied to understand the temperature effects involved in the mechanical stability of such clusters.

## 2. Computational methods

The starting geometries are those reported by [32] and then optimized at DFT level. All geometry optimizations were performed with B3LYP functional, which combines the exact Hartree-Fock exchange with the Lee, Yang and Parr correlation functional that includes the most important correlation effects [47]. All calculations were performed using the 6-31G(d) basis set [48,49]. This level of theory B3LYP/6-31G (d) has been widely used for the study of allotropic forms of carbon, successfully predicting the experimentally observed geometries [18–26,35]. Frequency calculations were carried out at the same level of theory in order to confirm that the optimized structures correspond to a local minimum.

In order to assess chemical stability, aromaticity of the quasi-fullerenes was computed by NICS methodology as given by Schleyer et al [50] in the different rings of the structures. Additionally, we present chemical potential μ [51] and chemical hardness η [52] to evaluate stability. This theory was employed with *the principle of chemical hardness*. We have used the following relations derived from DFT [53]:

$$\mu = \left(\frac{\partial E}{\partial N}\right)_{T,v(r)} \qquad (1)$$

and

$$\eta = \frac{1}{2}\left(\frac{\partial^2 E}{\partial N^2}\right)_{T,v(r)} \qquad (2)$$

where $E$ is the total energy, $N$ is the total number of electrons in the system, $T$ corresponds to temperature and $v(r)$ is the external potential. We applied the well-known finite-difference approximation to evaluate μ and η, considering that $E$ varies quadratically with respect to the number of electrons. Consequently, these parameters [54] may be given in orbital grounds as:

$$\mu = -\left(\frac{IP + AE}{2}\right) \quad (3)$$

$$\eta = \left(\frac{IP - AE}{2}\right) \quad (4)$$

where *IP* is the ionization potential and *AE* is the electron affinity. These two quantities can also be define as *IP= E(+) –E* and *AE=E- E(-)*, where *E(+)* is the system energy in cationic state and *E(-)* corresponds to the system energy in anion state.

In order to quantify the stability of the quasi-fullerenes compared to fullerenes, we calculated the formation energy per atom ($E_{form}$):

$$E_{form} = \frac{(E_{cluster} - nE_c)}{n} \quad (5)$$

where $E_{cluster}$ is the total energy of the quasi-fullerene ($C_n$), *n* is the number of carbon atoms in the cluster and $E_c$ is the energy of the isolated carbon atom. Charge calculations were carried out in accordance with the NBO method [55]. To understand the effect of temperature which quasi-fullerenes are subjected to, *ab initio* molecular dynamics (MD) with the Atom-Centered Density Matrix Propagation (ADMP) [56–58] method at B3LYP/6-31G(d) level theory was used. The minimum points in the electrostatic surface potential (ESP) were studied to find the structural stability of the quasi-fullerenes in a temperature ramp from $T_1$=800 K to $T_2$=1000 K. Semi-empirical AM1 [59] method was employed to analyze the behavior of carbon cluster with a longer time of 1000 fs. All calculations were performed with the Gaussian 09 code, Revision D [60].

## 3. Results and discussion

### 3.1 Structural description

In this work, we adopted the nomenclature *Symetry group*-$C_{n-q}$ (where n = 20, 42, 48 and 60) that describes the quasi-fullerenes (q) under study. The typical nomenclature for its most stable isomers *Symetry group*-$C_n$ (fullerenes) was also used. The study of carbon clusters at the theory level of B3LYP/6-31G(d) has been used from nonconventional isomers to conventional fullerenes, in the case of structures formed by $C_{36}$, $C_{40}$, $C_{46}$, $C_{48}$, $C_{62}$ and $C_{82}$ [34-39]. Therefore, in this work, this methodology was used to study the chemical properties of the proposed quasi-fullerenes. Full geometry optimization was carried out and the molecular structures of quasi-fullerenes ($C_{20-q}$, $C_{42-q}$, $C_{48-q}$ and $C_{60-q}$) are show in Fig. 1. The formation energies per atom are presented in Table 1. These values are negative, indicating that the clusters are thermodynamically stable. On the other hand, fullerenes used to compare energy stability are reported in the supplementary information (SI).

The first small carbon clusters $C_{3v}$-$C_{20-q}$, is a 20-carbon atoms system (Fig 1a). It is important to highlight that sp1+sp1 bond formation is present in this system, which corresponds to a curved symmetry that has been reported to be stable somewhere else [61]. Furthermore $C_{3v}$-$C_{20-q}$ presents mixed hybridization states of $sp^1+sp^1$ and $sp^1+sp^2$. These cluster do not present rings characteristic of quasi-fullerenes because they have semi-rings with $sp^1 + sp^1$ bonds, around the structure. The average bond length for the $sp^1 + sp^1$ atoms is 1.242 Å and 1.403A for the bond lengths with $sp^1 + sp^2$ hybridization (see Fig.1 **a**). The maximum angle of torsion for the atoms bound $sp^1+sp^1+sp^2$ is 157.6º. The closed structure of the $C_{3v}$-$C_{20-q}$ has a radius of r= 2.822 Å, which is larger than $I_h$-$C_{20}$ (2.08 Å) isomer [62].

The second structure under studied is $C_{3v}$-$C_{42-q}$, (see Fig.1b) with 6 atoms bonding with $sp^1+sp^1$ hybridization, 6 atoms with $sp^2+sp^1$ hybridization and 36 atoms with $sp^2+sp^2$ hybridization. This cluster presents the particularity to form three different rings in its geometry: 4 four-membered, 6 six-membered and 3 eight-membered rings. Maximum angle of distortion is 156.5º for the eight-membered rings, 117.6º for six-membered rings and 89.04º for four-membered rings. The $C_{3v}$-$C_{42-q}$, has a radius of r= 3.24 Å, this is larger than that compared to the more stable isomer $D_3$-$C_{42}$ (2.790 Å) [62], which has been studied in the formation of endohedral metallofullerenes by trapping a hexavalent metal such as W [63]. Due to its size and varied topology, $C_{3v}$-$C_{42-q}$ is an attractive molecule in the capture of large metal atoms for the formation of new compounds.

The structure $C_3$-$C_{48-q}$ (see Fig.1c) is modeled by adding six atoms in the base of cluster $C_{3v}$-$C_{42-q}$, with 24 atoms bonding $sp^1+sp^1$ (1.220 Å) and 24 atoms bonding with $sp^1+sp^2$ hybridization (1.438 Å) and only 18 of these atoms bond are $sp^2+sp^2$ (1.406 Å). This bonding configuration form different types of rings in the structure in such a way that they have 4 rings of hexagons and 3 octagon rings. The maximum angle of torsion is 146.2º for eight-membered rings and 120.0º for six-membered rings. The radius of the cluster $C_3$-$C_{48-q}$ is around 3.398 Å, which is larger than the most stable isomer $C_2$-$C_{48}$ (2.85 Å) [62]. Due to the large diameter of the $C_3$-$C_{48-q}$, it maybe possible that this clusters presents the capability of encapsulating small molecules. A major feature on closed carbon cages of 48 atoms is the rich variety of stable isomers that can be formed (199 isomers).

The $C_{3v}$-$C_{60-q}$ structure (see Fig. 1d) modeled by Belenkov and Shahova [32], presents a mixture of hybridized bonding $sp^1+sp^1$, $sp^1+sp^2$ and $sp^2+sp^2$, in its structure. This structure presents 36 atoms with $sp^1+sp^1$ hybridization (1.267 Å) and 24 with $sp^2+sp^2$ hybridization (1.421 Å), which represents a structure with 3 square, 3 hexagon and 3 octagon rings, respectively. The maximum torsional angle is 92.3º for four-membered rings, 115.6º for six-membered rings and 150.5º for eight-membered rings. In addition 4 atoms in a semi-ring with sp1+sp1 hybridization form a torsional angle of 153.6º. The radius of the cluster $C_{3v}$-$C_{60-q}$ corresponds to r = 4.101 Å, which is larger than that of the most stable and known isomer $I_h$-$C_{60}$ [62].

The closed cage form of the $C_{3v}$-$C_{60-q}$ system, is due to the $sp^1 + sp^1$ bonds present in its geometry. In order to understand such electronic structure, we performed geometry optimizations on the $C_{30}H_{14}$ isomers that are proposed with two different dispositions: linear (Geom_1) and curved (Geom_2). The geometries were optimized at the same level and they are shown in Fig. 2. The relative energy between both structures is 68.33 kcal/mol, corresponding to Geom_1 the smallest energy. This may be due to the favorable delocalization of electrons due to the planarity of the molecule, which is related to the aromatic value (as computed by NICS methodology reported on Table A.1 of SI) found on the three rings represented by frontier MOs (see Fig.2.A on SI). On the other hand, due to the symmetry and delocalization of electrons in the six-member rings, the Geom_2 has chemical stability, which is also related to the aromatic value on its three rings.

Taking into account the isosurfaces of frontier molecular orbitals Geom_2 depicted in Fig. 2.A of SI, the different membered rings present in the quasi-fullerenes under study, may present a significant contribution of electronic density, which favors the formation of this type of molecules. The quasi-fullerenes in this study show three rings of eight-membered around their geometry, whose conformation has been studied in a cyclooctatetraene, with non-planar geometry, similar to the rings reported in this work.

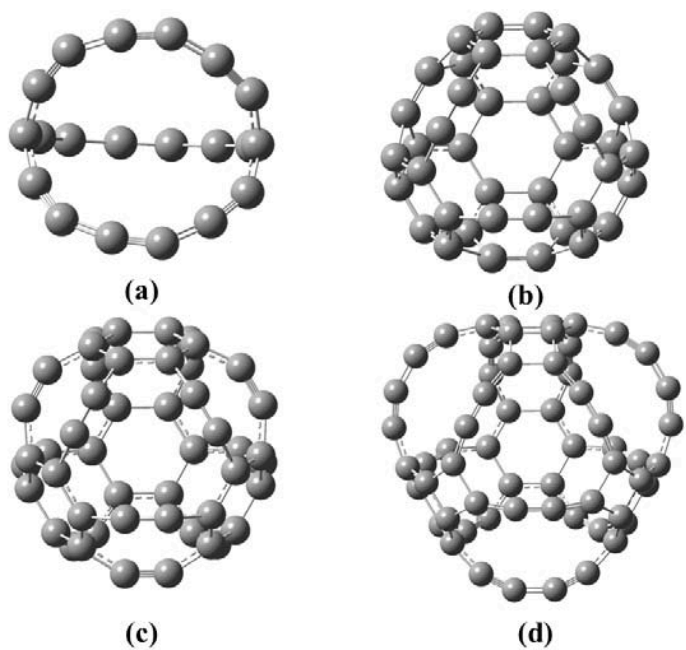

**Fig. 1 Optimized structures of (a) $C_{3v}$-$C_{20-q}$, (b) $C_{3v}$-$C_{42-q}$, (c) $C_3$-$C_{48-q}$ and (d) $C_{3v}$-$C_{60-q}$.**

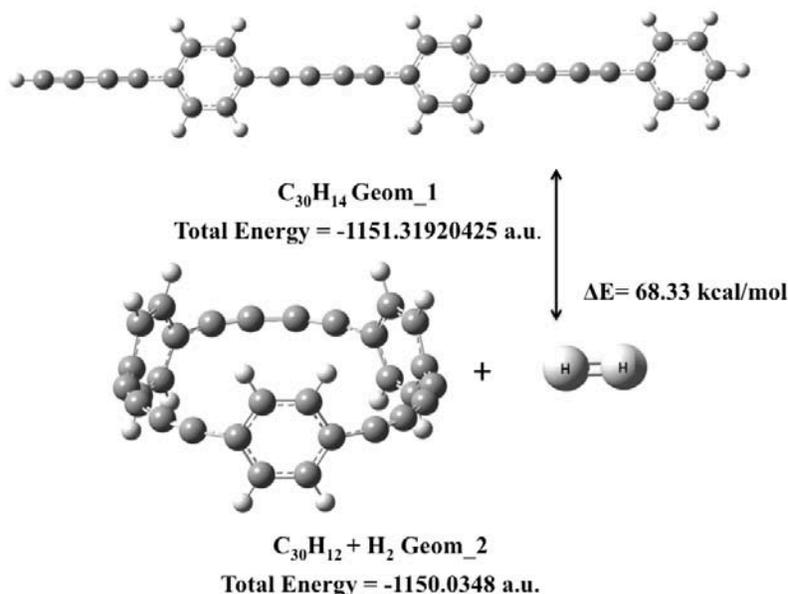

**Fig. 2** Structures optimized and Total Energy for $C_{30}H_{12}$ Geom_1 and $C_{30}H_{12}$ +$H_2$ Geom_2 with B3LYP/6-31G(d,p) level of theory.

**Table 1** Total energies of the clusters ($E_{cluster}$), radius of cluster (r), formation energies per atom ($E_{formation}$); calculated with B3LYP/6-31G(d) level of theory.

| Cluster | # Isomer | r (Å) | $E_{cluster}$ (a.u.) | $E_{formation}$ (Kcal/mol) |
|---|---|---|---|---|
| $C_{3v}$-$C_{20-q}$ | 0 | 2.82 | -761.42 | -185.24 |
| $I_h$-$C_{20}$ | 1 | 2.64 | -761.44 | -185.87 |
| $C_{3v}$-$C_{42-q}$ | 0 | 3.24 | -1599.50 | -192.95 |
| $D_3$-$C_{42}$ | 45 | 3.05 | -1599.96 | -199.78 |
| $C_3$-$C_{48-q}$ | 0 | 3.39 | -1827.85 | -190.95 |
| $C_2$-$C_{48}$ | 171 | 2.9 | -1828.65 | -201.4 |
| $C_{3v}$-$C_{60-q}$ | 0 | 4.10 | -2284.91 | -191.98 |
| $I_h$-$C_{60}$ | 1 | 3.55 | -2286.25 | -205.13 |

3.2 Chemical stability

Chemical stability of clusters has been studied according to the DFT, and also with semiempirical methods. The hardness is obtained from DFT parameters (see Eq. 4); nevertheless the ionization energy and electron affinity computed at the *ab initio* level are better than those at the DFT level, since it is know that, in calculations performed with DFT, the electronic density decays faster than at *ab initio* level, producing a gradient with higher density [64]. The hardness was computed from Eq. 4 along with the ionization energy and electron affinities calculated using Koopmans' theorem [53].

It was found that all clusters under study present high hardness values larger than 2 eV (see Table 2), except for $C_{3v}$-$C_{60\text{-}q}$ (1.87 eV), which has lowest chemical hardness. On the other hand, the $C_{3v}$-$C_{20\text{-}q}$ system presents a value of 3.24 eV, which corresponds to the most stable cluster of all quasi-fullerenes under study. Finally, the $C_{3v}$-$C_{42\text{-}q}$ and $C_3$-$C_{48\text{-}q}$ have similar hardness value, indicating high chemical stability of the systems.

On other hand, the energy gap ($\Delta E_{\text{LUMO-HOMO}}$) between the highest occupied molecular orbital (HOMO) and lowest unoccupied molecular orbital (LUMO) was also used as an indicator to determine energetic stability [54] of the quasi-fullerenes (see Table 2). In general, gaps larger than 1.3 eV indicate relative stability while gaps lower than 1.3 eV may indicate low stability [65]. The gap of quasi-fullerenes was determined at the ground state of each cluster. All gaps showed stability in all of the clusters. The $C_{3v}$-$C_{20\text{-}q}$ presents a large gap of 3.24 eV and the $C_{3v}$-$C_{60\text{-}q}$ system, presents a gap of 1.48 eV. We can see that all clusters (with the exception of $C_{3v}$-$C_{20\text{-}q}$) have properties of a semiconductor material, due to the gap size. The gaps of the $C_{3v}$-$C_{20\text{-}q}$, $C_{3v}$-$C_{42\text{-}q}$ and $C_3$-$C_{48\text{-}q}$ are larger than those of the most stable isomers ($I_h$-$C_{20}$, $D_3$-$C_{42}$ and $C_2$-$C_{48}$) with 1.95, 2.00 and 1.56 eV, respectively.

**Table 2 Ionization energy (EI), electron affinity (AE), chemical potential (µ), chemical hardness (η) and gap energies ($\Delta E_{\text{LUMO-HOMO}}$)**

| Clusters | EI (eV) | AE (eV) | µ (eV) | η (eV) | $\Delta E_{\text{LUMO-HOMO}}$ (eV) |
|---|---|---|---|---|---|
| $C_{3v}$-$C_{20\text{-}q}$ | 7.96 | 1.58 | -4.77 | 3.19 | 3.24 |
| $C_{3v}$-$C_{42\text{-}q}$ | 7.03 | 1.68 | -4.35 | 2.67 | 2.64 |
| $C_3$-$C_{48\text{-}q}$ | 7.1 | 2.04 | -4.57 | 2.53 | 2.52 |
| $C_{3v}$-$C_{60\text{-}q}$ | 6.56 | 2.82 | -4.69 | 1.87 | 1.48 |

3.3 Chemical reactivity

The reactivity of quasi-fullerenes has been studied using the Molecular Electrostatic Potential (MEP) isosurface [54]. Fig. 3 shows the MEP of all quasi-fullerenes presented in this study. Due to the topological variation present in the quasi-fullerenes, we analyzed the MEPs of different charge regions where the blue regions corresponds to sites of absence of electrons and red regions corresponds to sites of abundance of electrons. Therefore, to analyze of MEP in this type is important to locate sites that are attractive for a possible chemical attack.

The $C_{3v}$-$C_{20}$ has an isosurface the form of 3 semi-rings and positive charge regions are abundant in carbon sp$^2$ (see Fig. 3 a). Another main characteristic formed by the MEP is the region an absence of charge in the center of the molecule, which contradicts the negative value of the NICS indicative of delocalization of electrons in the center of the molecule. Discussed in the aromaticity section.

The $C_{3v}$-$C_{42}$ presents a single site with high concentration of electrons only one region without presence of any rings (see Fig. 3 b). On the other hand, in regions close to the six-membered rings regions of absence of electrons. For the 4 and 8 rings there is presence of regions of abundance electrons derived from the center of the molecule this site can be considered rich in electrons.

For the case $C_3$-$C_{48-q}$, the MEP presents of regions of abundant electrons are sites near the eight-membered rings (see Fig. 3 c). The regions with absence of electrons are given at sites close to six-membered rings. A particular characteristic with the MEP, are valleys formed in the regions of positive and negative charge, due to the topology of cluster.

As depicted in Fig. 3 d, the $C_{3v}$-$C_{60-q}$ shows regions of absence of electrons close regions to six-membered rings and high abundance in sites close to four-membered rings and at the top of molecule. The rich chemical reactivity and internal electronic density present in the quasi-fullerenes facilitates the interaction with charged systems for electronic charge storage. Because there are sites of donation and acceptance of electrons are presented around of the cluster. The different regions of absence and excess of electronic charge on the different ring membered in the quasi-fullerenes are due to the $\pi$ electrons distribution depicted in Fig. 5 for the frontier MOs.

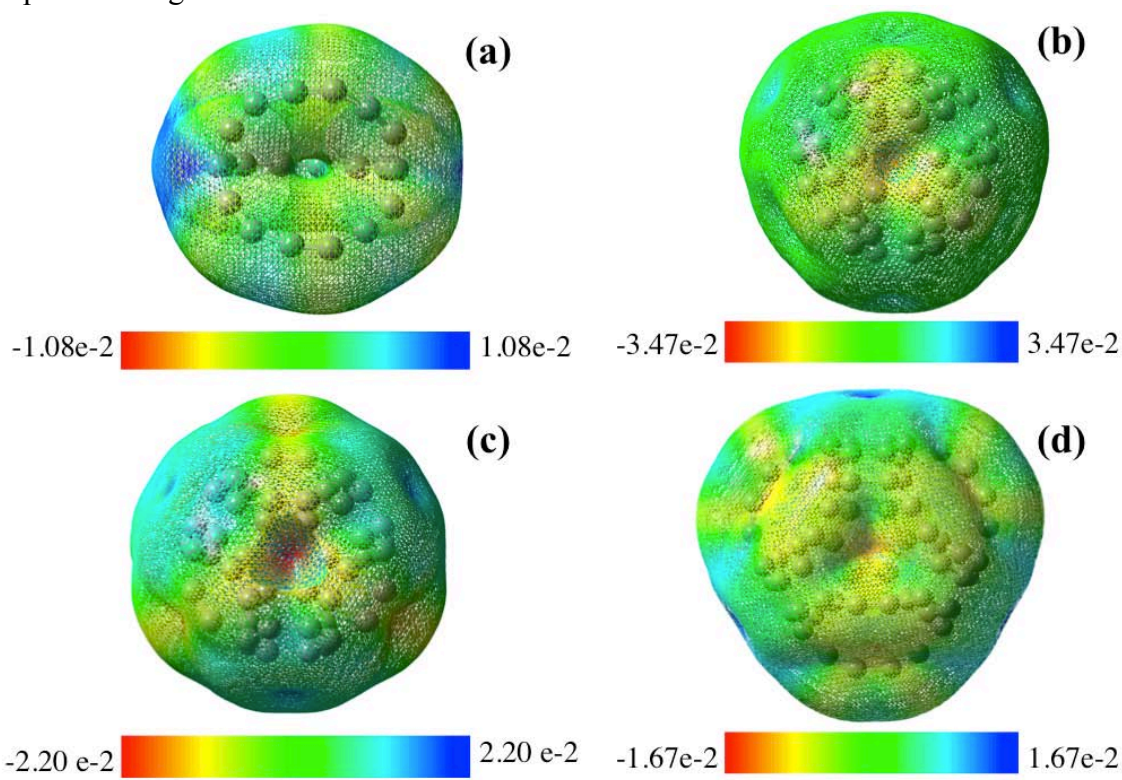

**Fig. 3 Mapping of the molecular electrostatic potential surface assigned to the electron density of (a) $C_{3v}$-$C_{20-q}$, (b) $C_{3v}$-$C_{42-q}$, (c) $C_3$-$C_{48-q}$ and (d) $C_{3v}$-$C_{60-q}$.**

3.4 Aromaticity

As it is well known, the closed-cage molecules have a strong aromatic character [66]. Whether a molecule is aromatic is well reflected by its NICS value. Magnetic shielding calculations using the NICS approach developed by Schleyer et al. [58] were carried out on the center of the quasi-fullerene cages and in the centers of the different ring members. Such values were compared with their isomers (fullerenes); the results are shown in Fig 4 and Table 3. The variation of different rings in quasi-fullerenes favors to found four-membered rings aromatics, which is unusual for systems that obey the Hückel rule.

In the case of $C_{3v}$-$C_{20-q}$, it appears to be a molecule with aromatic (-13.66 ppm) character compared to the lower energy isomer, despite the diameter difference of the fullerene $I_h$-$C_{20}$ (-18.78 ppm), the presence of electron delocalization in the geometric center suggests that $C_{3v}$-$C_{20-q}$ is polarizable cluster. To understand the aromatic character, we analyzed the frontier molecular orbitals (MOs) showed in Fig. 5 a, where HOMO and LUMO form molecular orbitals of π bonding around the molecule. Another criterion for studying the high values of magnetic susceptibility at the center of the quasi-fullerenes may be due to the degeneration (See Fig. 3-6.A of SI). The degeneration at the frontier MOs is indicative of the high delocalization of electrons in center cluster.

For the case of the $C_{3v}$-$C_{42-q}$ shows negative NICS values for all rings (4, 6 and 8-membered rings) which indicative an aromatic character in all their rings (see Fig. 4 b). The frontier MOs shows a majority contribution of molecular orbitals of π bonding close to aromatic rings as show in the Fig5 b. The high negative NICS values in the center of geometry is of character aromatic, otherwise for the $D_3$-$C_{42}$ (4.7 ppm) which reflects to be anti-aromatic molecule in the geometric center. The high values of the NICS in the center of $C_{3v}$-$C_{42-q}$ could be reflected by degeneration of frontier MOs like $C_{3v}$-$C_{20-q}$ case (See Fig. 4.A of SI).

The $C_3$-$C_{48-q}$ only presents two different type of 6 and 8 rings membered; both showed negative NICS values of -3.43 and -7.65 ppm respectively, indicative an aromatic character. The character aromatic can be interpreted by the MEP where red regions are close to the eight-membered where the sites of grater abundance of electrons are located. On the other hand, the negative NICS in the center of $C_3$-$C_{48-q}$ (see Table 3) is much smaller than its counterpart $C_2$-$C_{48}$ (-31.61 ppm). The frontier MOs show the highest contribution of molecular orbitals of π bonding around of eight-membered rings (See Fig. 5.A of SI) indicative of character aromatic in this ring.

For the case $C_{3v}$-$C_{60-q}$, the positive NICS values were reported for 6 and 8 rings membered, indicate the character anti-aromatic (see Fig. 4 d), these results are interpreted by the absence of electrons like showed with the MEP (see Fig. 3 d). Only the four-membered rings have negative NICS values (-5.26 ppm) that are reflected by the presence of frontier MOs (see Fig 5 d). The presences of π orbitals are shown around $sp^1$+$sp^1$ half-rings, whose electron stability on this bond was studied in the geometrical description section. The presence of anti-aromatic rings membered on this quasi-fullerene may be due to the absence of 6 and 8 rings membered and as a consequence this orbitals are located in the atoms that are around the structure. On the other hand, all the representations of the MOs of the quasi-fullerenes are located outside the molecules. This behavior manifests its reactive character to donate or give charge. The degenerations of MOs can be consulting in Fig.3-6.A of SI.

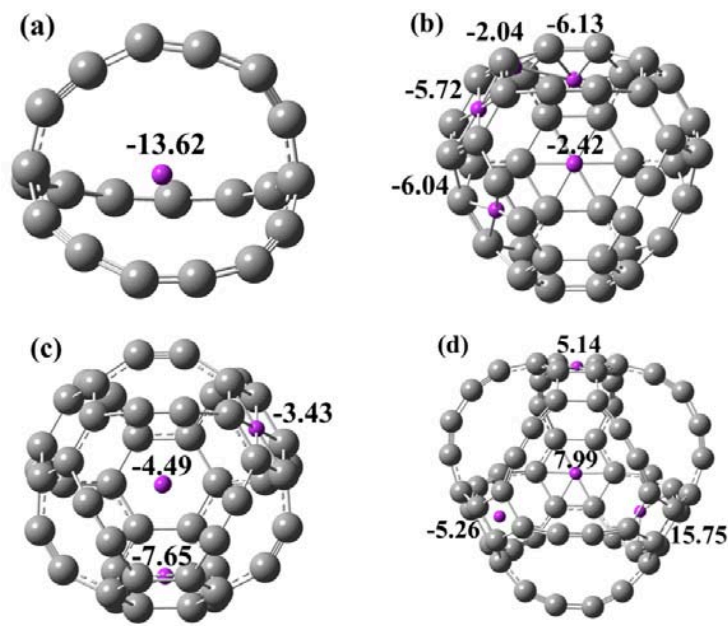

**Fig. 4 NICS values (ppm) at the center ring for of (a) $C_{3v}$-$C_{20-q}$, (b) $C_{3v}$-$C_{42-q}$, (c) $C_{3}$-$C_{48-q}$ and (d) $C_{3v}$-$C_{60-q}$, were computed at the B3LYP/6-31G(d) level.**

The spatial distribution of MOs on clusters $C_{3v}$-$C_{20-q}$, $C_{3v}$-$C_{42-q}$ and $C_{3}$-$C_{48-q}$ are shown in the Fig. 5. The distribution of HOMO-LUMO in the quasi-fullerenes around the aromatic ring membered can be observed. Bonding MOs are clearly observed on the aromatic ring members. In conclusion the aromaticity in this type systems are interpreted through the contribution of frontier MOs where the aromatic sites form molecular orbitals π bond.

**Table 3 Value of NICS in center of geometry with B3LYP/6-31G(d)**

| Cluster | NICS (ppm) |
|---|---|
| $C_{3v}$-$C_{20-q}$ | -13.66 |
| $I_{h}$-$C_{20}$ | -18.78 |
| $C_{3v}$-$C_{42-q}$ | -10.55 |
| $D_{3}$-$C_{42}$ | 4.7 |
| $C_{3}$-$C_{48-q}$ | -4.48 |
| $C_{2}$-$C_{48}$ | -31.61 |
| $C_{3v}$-$C_{60-q}$ | 14.73 |
| $I_{h}$-$C_{60}$ | -2.69 |

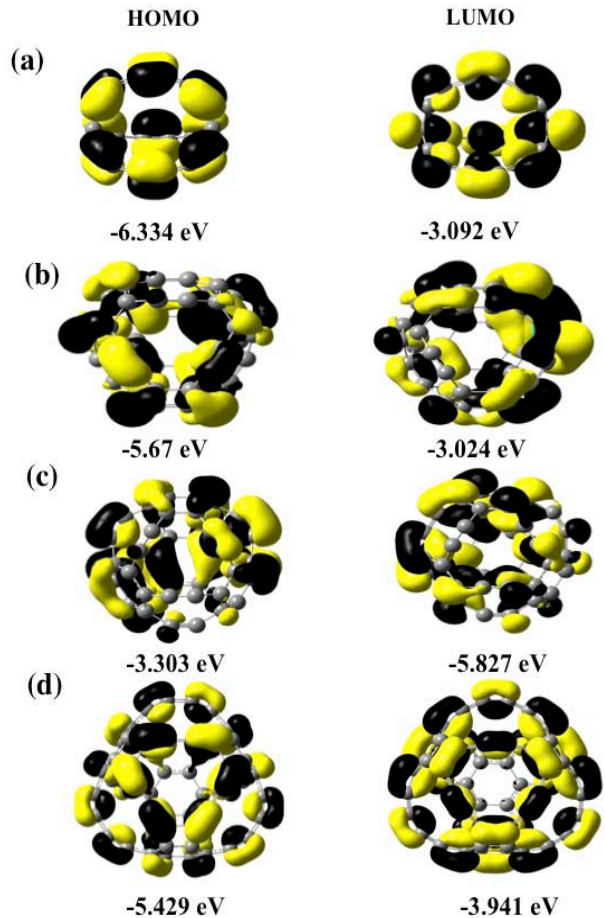

**Fig. 5** Spatial representations of frontier MOs on the quasi-fullerenes under study (a) $C_{3v}$-$C_{20\text{-}q}$, (b) $C_{3v}$-$C_{42\text{-}q}$, (c) $C_3$-$C_{48\text{-}q}$ and (d) $C_{3v}$-$C_{60\text{-}q}$.

3.5 Thermal Stability

To study the thermal stability of quasi-fullerenes we employed the Atom-Center density matrix propagation (ADMP) [56–58] method to understand the potential energy surface (PES) of this clusters, we employed a temperature ramp for all proposed systems with an initial temperature from $T_1$=800 K to $T_2$=1000 K. We chose this temperature regime since was taken by the method of FTCP pyrolysis [28]. In Fig. 6 we showed the ADMP molecular dynamics profile of the PES, which was calculated with 100 steps of 1 fs.

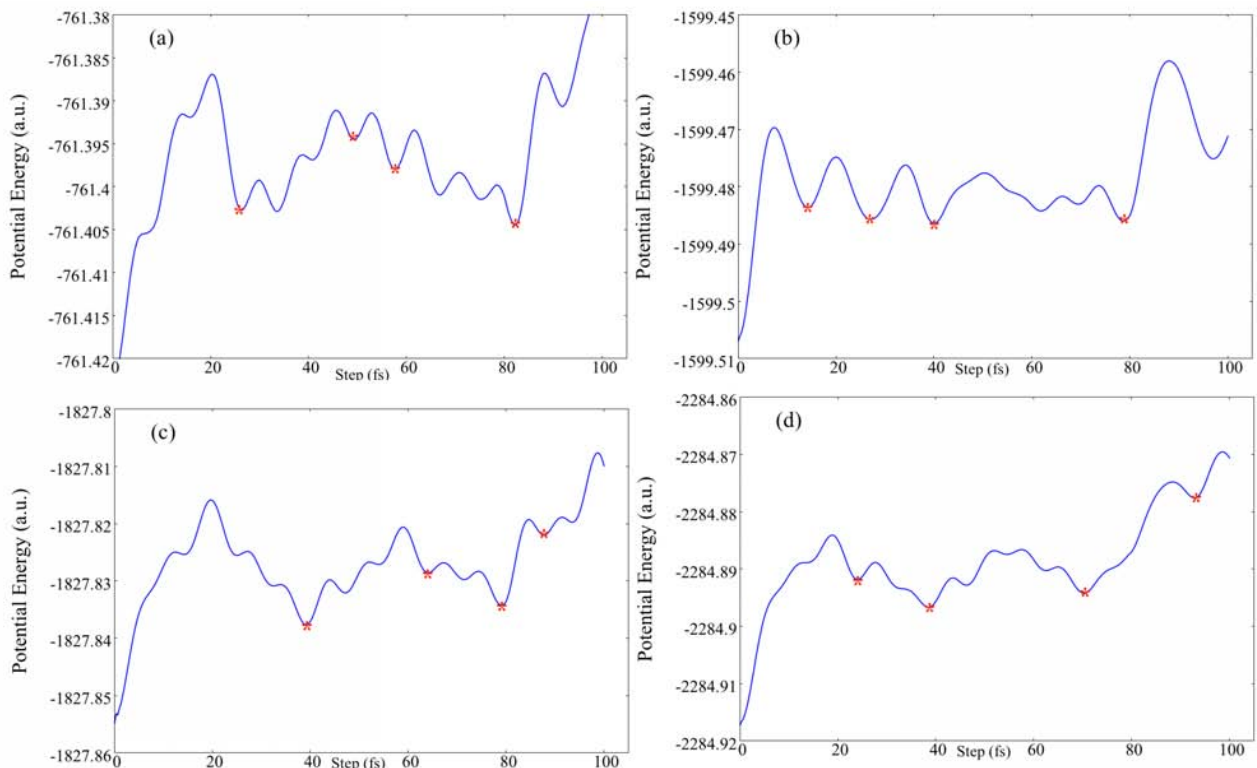

**Fig. 6 ADMP molecular dynamics profile of the potential energy surface (PES) computed with B3LYP/6-31(d) level for (a) $C_{3v}$-$C_{20\text{-}q}$, (b) $C_{3v}$-$C_{42\text{-}q}$, (c) $C_3$-$C_{48\text{-}q}$ and (d) $C_{3v}$-$C_{60\text{-}q}$.**

The MD simulation started from ground state structure previously found at B3LYP/6-31G(d) level. We found the potential energy profile with a variation from 0.05 to 0.1 a.u., which shows slight modifications in the bond lengths of the clusters. In the $C_{3v}$-$C_{20\text{-}q}$ cluster the average minima bond length is 1.24 Å and a maximum is 1.36 Å. For the case of $C_{3v}$-$C_{42\text{-}q}$ and $C_3$-$C_{48\text{-}q}$ the average bond lengths are similarly with 1.38 Å and 1.40 Å for the minima and 1.48 Å the greatest bond length. The $C_{3v}$-$C_{60\text{-}q}$ cluster exhibits a bigger variation in its bond lengths during the simulation with an average minima bond length of 1.37 Å and a maximum of 1.53 Å. Despite the high temperature values at which the carbon clusters are subjected, the distortion of the different ring membered is evident in the peaks of the PES, however there is no breaking of a ring or conformation of some new ring in the structure of the quasi-fullerenes. Obviously the geometries of the quasi-fullerenes lose the symmetry of the ground state. We further re-optimized two of the minima found at the PES (see red asterisk in Fig.6) at B3LYP/6-31G(d) level of theory, which correspond to the geometries in grounds state previously mentioned. The effect of temperature on these carbon clusters is of great relevance in the formation of this type of molecules. This information maybe directly related to the experimental synthesis route of new carbon nanostructures.

An additional study of thermal stability to that described above, was carried out with semi-empirical AM1 method, with the purpose to understand the evolution of the system subjected to the same temperature regime with a longer time of 1000 fs, reported in Fig. 7.

The results showed structural stability on the carbon cluster under study. In all cases the interval values between the maximum and minimum has an average variation of two numbers significant (of 0.19 a.u. on average), indicative of high thermal stability. The rings of all quasi-fullerenes are distorted during the simulations but no bond breaking was reported.

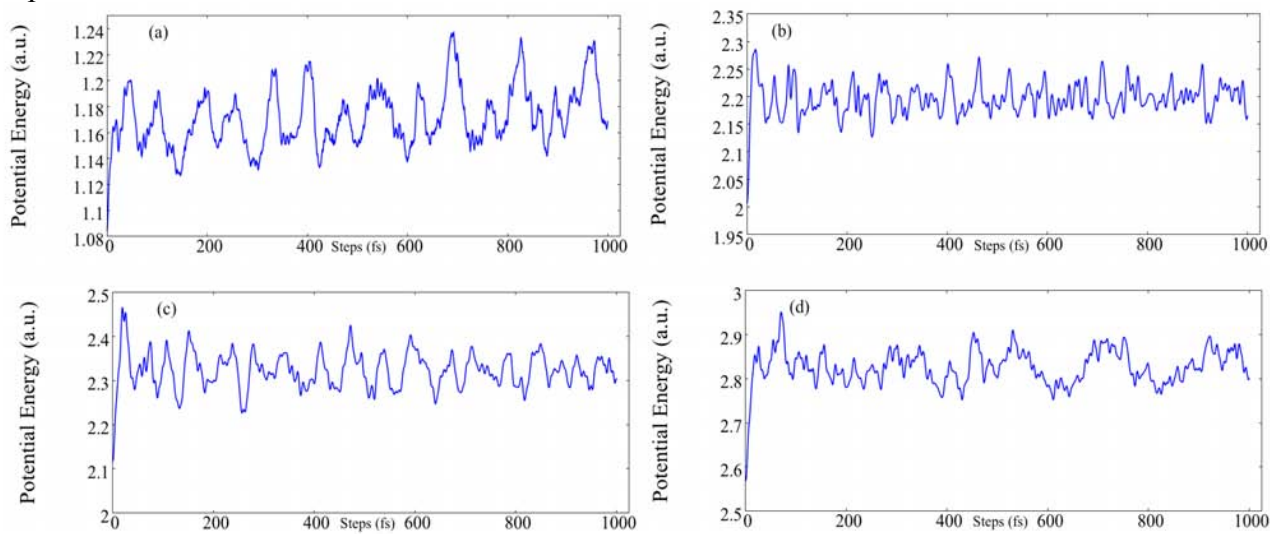

**Fig. 7** ADMP molecular dynamics profile of the potential energy surface (PES) calculated with AM1 method, (a) $C_{3v}$-$C_{20\text{-}q}$, (b) $C_{3v}$-$C_{42\text{-}q}$, (c) $C_3$-$C_{48\text{-}q}$ and (d) $C_{3v}$-$C_{60\text{-}q}$.

**4. Conclusions**

We have studied the electronic structure and stability of quasi-fullerenes $C_{n\text{-}q}$, with n=20, 42, 48, 60 using DFT. We studied new allotropic forms of closed cage carbon, which have a high chemical stability than that of the more stable isomers (fullerenes). The aromatic character in the different ring incorporated in the quasi-fullerenes, describes the favorable electron delocalization in this type of molecules, where the $C_{3v}$-$C_{42\text{-}q}$ is the most aromatic cluster. On the other hand, the study of MEPs was used to analyze chemical reactivity, showing different and attractive sites of higher electron density, favorable for charge transfer interactions. The energy gaps (HOMO-LUMO) are used as an indicator for the kinetic stability of carbon isomers. The molecular dynamics calculations showed the thermal stability of the quasi-fullerenes when subjected to high temperatures (800 K - 1000 K), this study confirms structural stability of quasi-fullerenes. This group of quasi-fullerenes presents interesting properties from chemical reactivity to a greater diameter in their structural cavities, which are attractive for promising applications such as energy storage, drug delivery, among others.

**Acknowledgements**

Universidad Nacional Autónoma de México (UNAM) for the computer resources, DGAPA and CONACYT for financial support Ph.D. Scholarship No. 539402. (Christian A. Celaya). J.M. wants to acknowledge the support given by Cátedras-CONACYT (Consejo Nacional de Ciencia y Tecnología) under Project No. 1191; the sup- port given by CONACYT through Project SEP- Ciencia Básica No. 156591; DGTIC (Dirección General de Cómputo

# Supplementary Information

# NEW NANOSTRUCTURES OF CARBON: QUASI-FULLERENES $C_{n-q}$ (n= 20, 42, 48, 60)

Christian A. Celaya[1], Jesús Muñiz[2,3], Luis Enrique Sansores[1*]
[1]Departamento de Materiales de Baja Dimensionalidad, Instituto de Investigaciones en Materiales, Universidad Nacional Autónoma de México, Apartado, Postal 70-360, Ciudad de México, 04510, México.
[2]Instituto de Energías Renovables, Universidad Nacional Autónoma de México, Priv. Xochicalco s/n, Col. Centro, Temixco, Morelos CP 62580, México
[3]CONACYT-Universidad Nacional Autónoma de México, Priv. Xochicalco s/n, Col. Centro, Temixco, Morelos CP 62580, México

Stability Chemical of Classics Fullerenes

The stables isomers were optimized with B3LYP/6-31G(d) level theory.

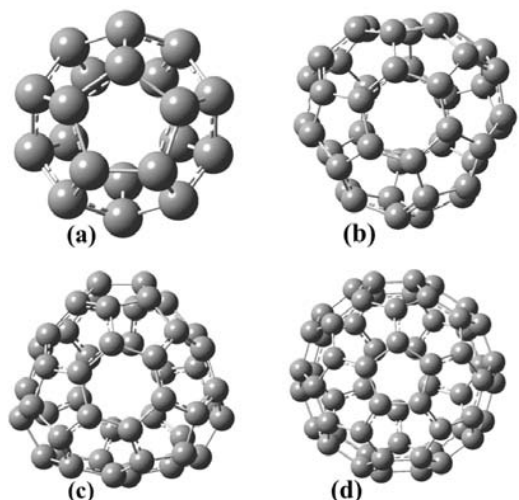

**Fig. 1A** Optimized structures of (a) $I_h$-$C_{20}$, (b) $D_3$-$C_{42}$, (c) $C_2$-$C_{48}$ and (d) $I_h$-$C_{60}$.

Chemical Stability

**Table 1.A** NICS values reported with B3LYP/6-31G(d,p) level.

| System | Ring 1 | Ring 2 | Ring 3 | Center |
|---|---|---|---|---|
| Geom_1 | -9.46 ppm | -9.06 ppm | -9.25 ppm | --- |
| Geom_2 | -6.96 ppm | -6.96 ppm | -6.97 ppm | 2.39 ppm |

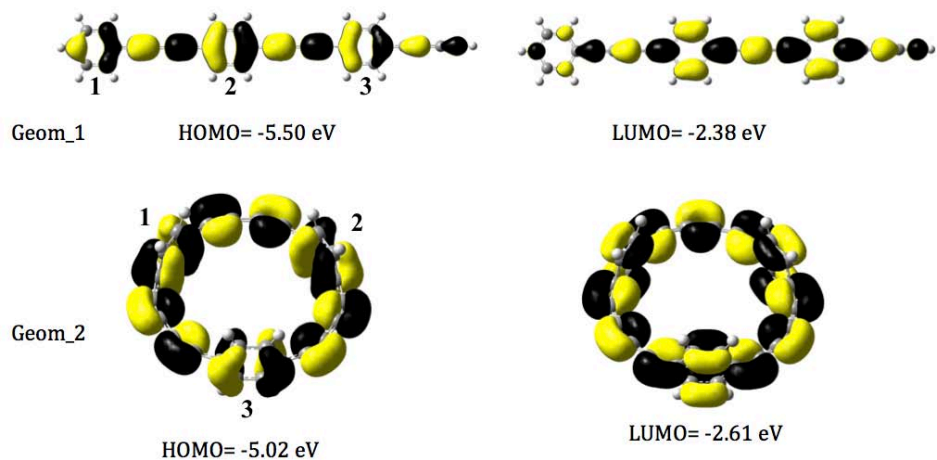

**Fig. 2.A** Spatial representations of frontier MOs of Geom_1 and Geom_2.

Spatial representation of frontier MOs of Quasi-Fullerenes under study.

| eV | MOs |
|---|---|
| (-6.34) | HOMO |
| (-6.44) | HOMO-1 |
| (-6.69) | HOMO-2 |
| (-6.69) | HOMO-2 |

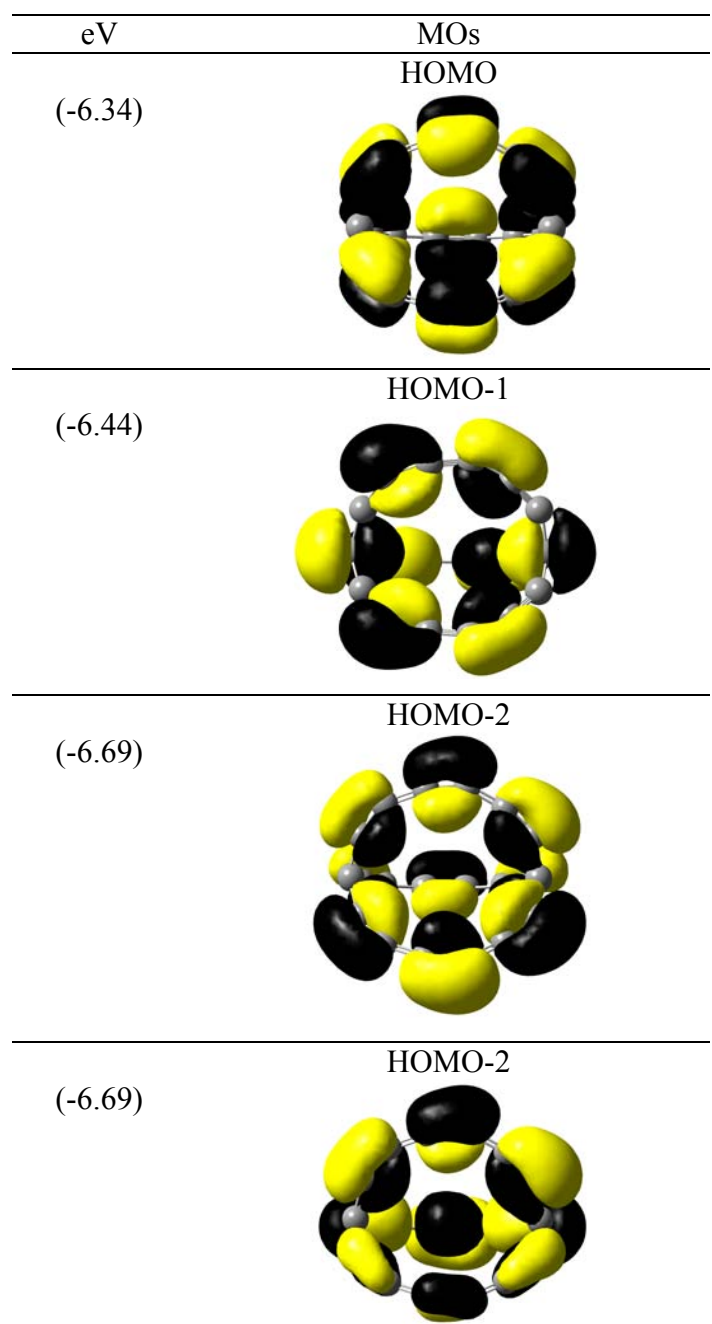

**Fig. 3.A** Spatial representations of MOs HOMO, HOMO-1 and HOMO-2 for $C_{3v}$-$C_{20\text{-}q}$.

| eV | MOs |
|---|---|
| (-5.67) | HOMO |
| (-5.67) | HOMO |
| (-6.221) | HOMO-1 |
| (-6.228) | HOMO-1 |

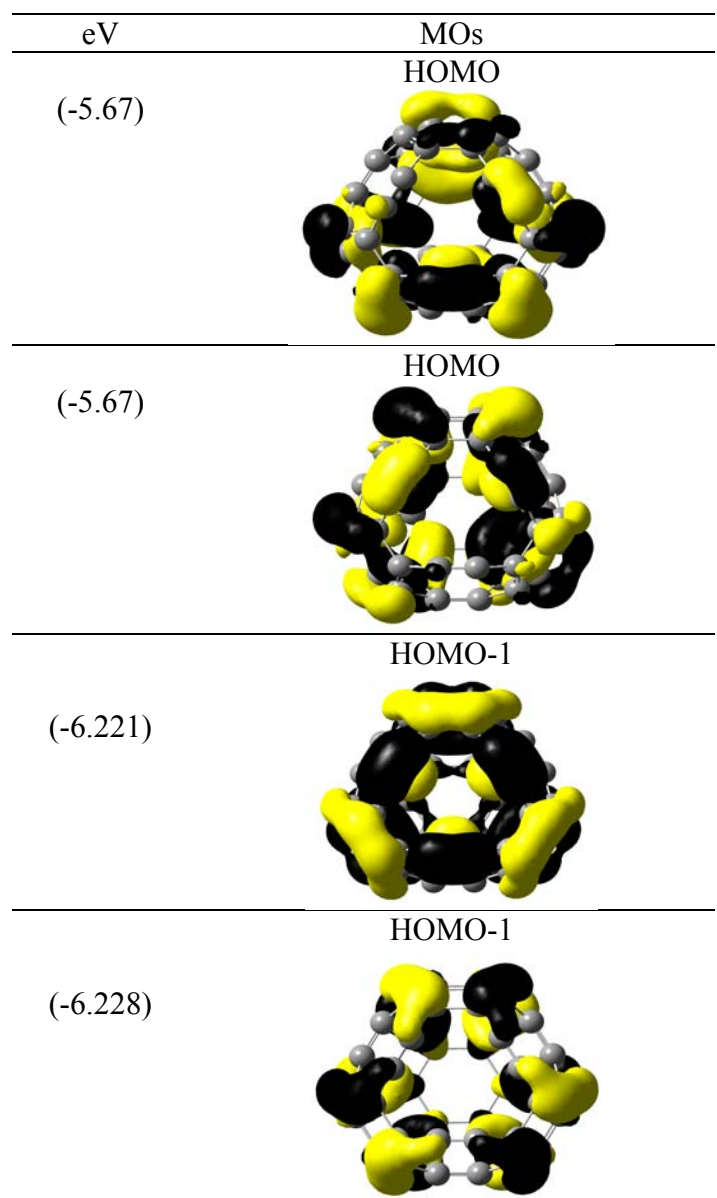

**Fig. 4.A** Spatial representations of MOs HOMO and HOMO-1 of $C_{3v}$-$C_{42\text{-}q}$.

| eV | MOs |
|---|---|
| (-5.827) | HOMO 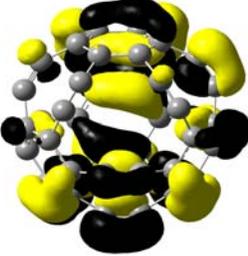 |
| (-5.827) | HOMO 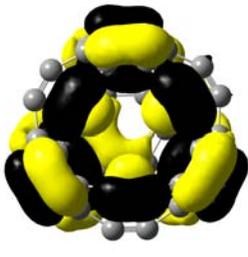 |
| (-5.828) | HOMO 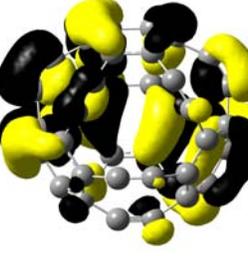 |
| (-6.589) | HOMO-1 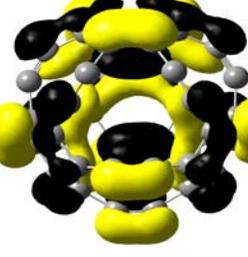 |

**Fig. 5.A** Spatial representations of MOs HOMO and HOMO-1 of $C_3$-$C_{48-q}$.

| eV | MOs |
|---|---|
| (−5.429) | HOMO 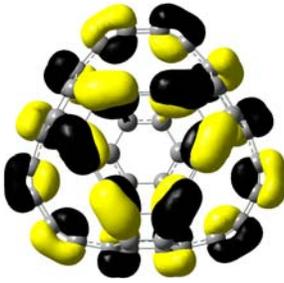 |
| (−5.645) | HOMO-1 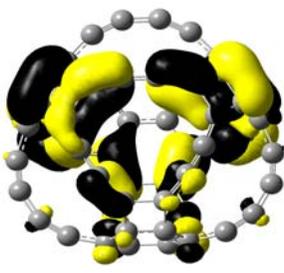 |
| (−5.646) | HOMO-1 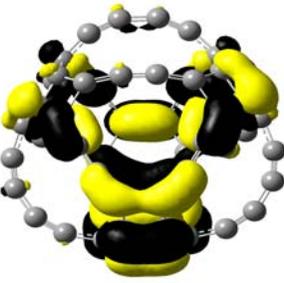 |
| (−6.375) | HOMO-2 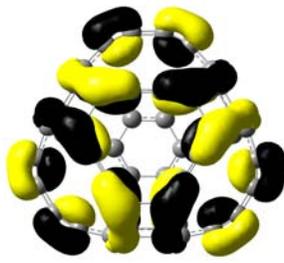 |

**Fig. 6.A** Spatial representations of MOs HOMO, HOMO-1 and HOMO-2 of $C_{3v}$-$C_{60\text{-q}}$.